\documentstyle[seceq,preprint]{ptptex}
\def\eq{\begin{equation}} 
\def\en{\end{equation}} 
\newcommand{\beqn}{\begin{eqnarray}} 
\newcommand{\eeqn}{\end{eqnarray}} 
\title{%        %You can use \\ for explicit line-break
Alternative to Higgs and Unification
}
\subtitle{renormalization and essential singularity} 

\author{%       %Use \sc for the family name
Miyuki {\sc Nishikawa}\footnote{E-mail: nisikawa@hep-th.phys.s.u-tokyo.ac.jp}
}

\inst{%         %Affiliation, neglected when [addenda] or [errata]
Physics Department, University of Tokyo, Tokyo 113
}
\recdate{%      %Editorial Office will fill in this.
}

\abst{%       %this abstract is neglected when [addenda] or [errata]
 In this paper, we discuss the self-consistency condition for the spherical 
symmetric Klein-Gordon equation, and then discuss a natural possibility
 that gravity and weak coupling constants $g_G$ and $g_W$ may be defined 
after $g_{EM}$. In this point of view, gravity and the weak force are 
subsidiary derived from electricity. Particularly, $SU(2)_L\times U(1)$ 
unification is derived without assuming a phase transition. A possible 
origin of the Higgs mechanism is proposed. Each particle pair of the 
standard model is associated with the corresponding asymptotic expansion 
of an eigen function.
}

\begin{document} 
\maketitle
\section{Introduction } 
 In usual dimensional counting, momentum has dimension one. But 
a function $f(x)$, when differentiated $n$ times,  
does not always behave like one with its power smaller by $n$. 
For example, this can occur in the neighborhood of $x=0$ 
if the function $f(x)$ has an essential singularity at $x=0, 
f(0)\to 0 (x\to 0)$. Thus a dimension of momentum is such an 
$\it operater$ that cannot be fixed unless the operand of the 
differential operator is explicit. This inevitable uncertainty 
may be essential in general theory of renormalization, including 
quantum gravity. 

As an example, we first consider a one-dimensional 
Schr\"{o}dinger equation, noting the two possible cases of 
singularities for a potential in section 2. In section 3
we construct the most general type of a singularity that is closed in 
usual operations. Then we classify the singularities 
for a potential by comparing the eigen function $y(x)$ with its second 
derivative, assuming that $y$ is $C^2$-class. The result crucially 
depends on the analytic property of the eigen function near its 0 point. 
In cases the eigen function has an essential singularity at $z=0$ 
which is $C^2$-class when approached from the positive real axis, 
the eigen function does not satisfy the Lipschitz continuity condition. 
Then the solution is not always unique, which may be the origin of 
the gauge ambiguity important to quantum field theory. Section 4 is 
devoted to a philosophical discussion of this kind.

Section 5 is the extension to higher dimensions and the long distance limit. 
We consider the Klein-Gordon equation with a spherical symmetric 
$U(1)$ potential $A^\mu :=(\phi(r), 0, 0, 0)$, assuming that 
the potential $\phi$ has at least one normalizable eigen function 
$R(r)$, which in turn creates another potential $\phi '\propto R$. 
There are 10 cases in the long distance limit. 

Applying the results of the previous section, we can obtain 
theorems for the long distance limit in section 6. 
Particularly, we can prove $SU(2)_L\times U(1)$ unification without 
assuming a phase transition or the detailed Higgs mechanism. 
Section 7 is a proposal for the origin of the Higgs mechanism, 
making use of the asymptotic mathematical ambiguity of the expansions 
or degenerate eigen functions. In section 8 
we try to deal with gravity within the framework of the standard model 
$+\alpha$ and redefine gravity as an integral constant of the laplacian. 
Furthermore, we associate each pair of the particles in the 
standard model with the corresponding asymptotic expansion. 
\section{Possibility of singularity and domain of definition } 
 For simplicity, let us first consider a one-dimensional 
Schr\"{o}dinger equation 
\eq -y'' +V(x)y = Ey. \label{1} \en 
$y(x)$ is defined in $0<x<x_0$, where $x_0$ is some positive constant.
In fact, any $C^2$-class function $y(x)$ satisfies (\ref{1}) if  
we take 
\eq V=y''/y, \;\; E=0. \label{ko2 } \en 
Here the replacement of the constant $E\to E'$ is equivalent to  
$e\phi \to e\phi -E'+E$, so from now on we take $E=0$. There are 2 possible  
cases for (\ref{ko2 }) to have a singularity at $x\to +0$: 
 
(I) $y(x)\to 0$ for $x\to +0$,  
 
(II) $y''$ does not converge for $x\to +0$. 

For (I), the $V(x)$ in the L.H.S. of (\ref{ko2 }) is called
 a potential iff there exists at least one $C^2$-class eigen function
$y(x)$ satisfying (\ref{1}). 
\section{Classification of the power of possible singularities } 
 Now let us move the possible singularity to $x=0$ by the 
redefinition of the origin and consider the behavior of $V$ as  
$x\to +0$. Let $y(z)$ be the natural analytic continuation of $y(x)$ 
(from the real axis) to the complex plane.
 
\vspace{.2 in} 
\noindent (CASE 1) $y(z)$ has no essential singularity at $z=0$. 
 
\vspace{.1 in} 
\noindent (a) If $y(z)$ can be Laurent expanded around $z=0$ as  
\beqn y=\sum_{n=k }^\infty a_nz^n, \;\; a_k\neq 0, \label{ko4 }  
\eeqn 
then  
\beqn \frac{y'' }{y }=\frac{\sum_{n=k }^\infty  
a_nn(n-1)z^{n-2 }}{\sum_{n=k }^\infty a_nz^n} \to\left  
\{ \begin{array}{l}\frac{a_d }{a_k }d(d-1)z^{d-2-k }\;\; (0\leq  
k)  \\ 
k(k-1)z^{-2 }\;\; (k<0) \end{array} \right. , \label{ko5 } \eeqn 
with $d$ the lowest power such that $a_d\neq 0$ and $1<d$ (if  
there is no such $d$, $a_d=0$). 
 
\vspace{.1 in} 
\noindent (b) When we replace the power of the finite number of  
terms in the type (a) expansion with an arbitrary real number,  
{\footnote{From now on, the expansion coefficients are all real  
except if mentioned, and the branch is chosen so that the 
function takes unique real value at $z\to +0$. More precisely, a  
branching point with the power of an irrational number is an  
essential singularity, but the difference is not important  
here\cite{uniq}.}}  
\beqn \frac{y'' }{y }\to\left\{ \begin{array}{l}\frac{a_d }{a_k } 
d(d-1)z^{d-2-k 
}\;\;\left (\begin{array}{l}\mbox{if}\;  y = a_0 +a_1z +a_dz^d\cdots  
\;\; \mbox{or} \\ 
y = a_1z +z_dz^d\cdots \end{array} \right )\\ 
k(k-1)z^{-2 }\;\; (k<0)\;\; \mbox{(otherwise)} \end{array} \right. , 
\label{ko6 } 
\eeqn 
where $a_d$ is the coefficient of the lowest power except for  
$0, 1$. 
Thus far, the powers $\nu$ where the potential can  behave like  
$V\to x^\nu$ as $x\to +0$ are 
 
 for (I), $\nu =-2 \; ;\; -1\leq \nu$,  
 
 for (II), $-2\leq \nu <0$. 
 
\noindent (c)  
\beqn y=\sum_{n=l }^ka_n(\log z)^n, \;\; a_l\neq 0, \label{ko7 }  
\eeqn 
if the above expansion is possible, then 
\beqn \frac{y'' }{y }=\frac{\sum_{n=l }^kna_n\{ (n-1)(\log z)^{n- 
2 }-(\log  
z)^{n-1 }\} }{z^2\sum_{n=l }^ka_n(\log z)^n }\to\frac{-k }{z^2 
\log z } \label{ko8 } \;\; ,  
\eeqn 
where $\log z$ diverges as $z\to 0$, but for an arbitrary integer  
$n$, $z(\log z)^n$ tends to $0$. So we can regard $\log z$ as `an infinitely  
small negative power' $z^{-\epsilon }\;(\epsilon >0)$. Then we  
can generalize type (b) expansion by the replacement of the  
finite number of terms 
\beqn a_nz^n\to z^n\sum_{m=l_n }^{k_n }a_{mn }(\log z)^m\;\;  
(m\in R). \label{ko9 }  
\eeqn 
This has the effect of 
\beqn \left \{ \begin{array}{l}z^{d-2-k }\to z^{d-2-k }(\log z) 
^m\;\; 
       (m\in R)\\ z^{-2 }\to z^{-2 }/\log z \end{array} 
\right. \label{ko10 } \eeqn 
in (\ref{ko6 }), i.e.,  
\beqn \mbox{for (I), } \nu =-2 (+\epsilon )\; ;\;\; -1\leq \nu . 
\label{ko11 }  
\eeqn 
{\footnote{For a $C^2$-class function $y$, $-1-\epsilon$ is  
impossible. And for (II), the region of $\nu$ is invariant.}}  
Let us call this type of expansion type (c). For type (c) expansions, 
We can define the index of power $k_y, \mu_y, \nu_y (z\to +0)$ 
as follows: \beqn y\to z^{k_y }, \;\;\frac{y' }{y }\to 
z^{\mu_y }, \;\;\frac {y'' }{y }\to  
z^{\nu_y }.  \label{ko12 } \eeqn 
Type (c) property is invariant under finite times of  
summations, subtractions, and differentiations. 
  
\vspace{.1 in} 
\noindent (d) When we apply finite times of summations,  
subtractions, multiplications, divisions (by $\neq 0$),  
differentiations, and compositions (with the shape of $f(g(z)), \; 0 
\leq k_g, \; g( +0)= +0$ where $f, g$ are type (c) expansions),  
$k_y, \mu_y, \nu_y$ can also be defined. As an arbitrary type (d)  
expansion $f(z)$ has a countable number of terms and a  
nonzero `radius of convergence 
{\footnote{The meaning of this term is different from the usual one 
because $z=0$ can be a singularity point.}}  
' $r$ where the expansion  
converges for $0<|z|<r$, it can be written as  
\beqn f(z)=\sum_{n=0 }^\infty f_n \;\; . \label{ko13 } \eeqn 
As the `principal part' which satisfies $k_{f_n}<0$ consists  
of finite number of terms, a type (d) expansion diverges or  
converges monotonically as $z\to +0$, so enables  the expansion of  
(\ref{ko13 }) in the order of ascending powers. As the expansion  
is almost the same as that of type (c) (the only differences are the  
multiplications by $(\log z)^n$ for an infinite number of terms  
and the appearance of the terms like $\log (z\log z)$), the  
region of $\nu_y$ remains.  
 
\vspace{.2 in} 
\noindent (CASE 2) $y(z)$ has an isolated essential  
singularity at $z=0$. 
In complex analysis, a sequence of points can converge to any  
value depending on its approach to an essential singularity (with  
infinite order) \cite{Ahlf }. But now that we deal with only the  
case along the real axis $z\to +0$, the limit is sometimes well  
defined. Let us study the following cases.  
 
\vspace{.1 in} 
\noindent (e) When the following expansion is possible (type (e)): 
$y=\pm e^{f(z) }$, where $f$ is a type (d) expansion. We can  
define the finite values $\mu_y, \nu_y$ by  
\beqn \left \{ \begin{array}{l}\frac{y' }{y }=f'\to z^{\mu_y },  
\;\; \mu_y=k_f +\mu_f , 
\nonumber \\ 
\frac{y'' }{y }=f'^2 +f''\to z^{\nu_y }, \;\;\nu_y\geq {\mbox{min}}(2k_f  
+2\mu_f, \;  
k_f +\nu_f) .\end{array} \right. \label{ko14 } \eeqn 
Let us consider the region of $\nu_y$. For $k_f\geq  
0$ it is the same as for the type (d). For  
\beqn y=e^{az^k }, \;\; a, \; k\in R, \;\; k\leq 0 \label{ko15 }  
\eeqn 
satisfies 
\beqn  
\frac{y'' }{y }=a^2k^2z^{2k-2 } +ak(k-1)z^{k-2 }\to\left\{  
\begin{array}{l} 
z^{-2\pm\epsilon }\;\;(k=-\epsilon )\nonumber \\ 
a^2k^2z^{2k-2 }\;\; (k<0) \end{array}, \right. \label{ko16 }  
\eeqn 
combination with type (c) case leads to the region  
of $\nu_y$ being:  
 
for (I), $\nu_y\leq -2 +\epsilon \; ; \;\; -1\leq \nu_y$, 
 
for (II), an arbitrary negative number.  
 
\noindent Let us then consider if we can fill the remaining `window'  
of the region of $\nu_y$ for (I),  
 
$-2 +\epsilon <\nu_y<-1$. 
 
\vspace{.1 in} 
\noindent (f) When we can write $y=f_0 +\sum_{n=1 }^{m }(\pm )e^ 
{f_n }$, where $f_n$ is of type (c), $k_{f_n }<0$, and ($\pm$)  
takes each of the signatures $ +-$. 
We can assume that each terms in $\sum$ are ordered in the  
increasing absolute values for $z\to +0$. Because 
\beqn e^{az^k }\to\left \{\begin{array}{l}z^0\;\;(k\geq 0, \;  
a\neq 0)\nonumber \\ 
0\;\;\; (k<0, \; a<0)\nonumber \\ 
\infty\;\; (k<0, \; a>0) \end{array}\right. \label{ko18 } \eeqn  
and $y\to 0$ for (I),  
 
\beqn y=\left (\sum_{n=0 }^\infty\sum_{m=l_n }^{m_n }a_{nm }z^n 
(\log  
z)^m\right ) +\sum_{n=1 }^l(\pm ) 
e^{\sum_{i=k_n }^\infty\sum_{j=l_{ni } }^{k_{ni } }a_{nij }z^i(\log z)^j 
} 
. \label 
{ko19 } \eeqn  
If the second term sum at the R.H.S. is not 0, we can write 
 
\beqn k_l<\cdots <k_1<0, \;\; a_{nk_nk_{ni } }<0. \label{ko20 }  
\eeqn  
As $y$ is of $C^2$-class, the first term can be written  
as  
\beqn (\;\; )=a_{10 }z +\sum_{n=2 }^\infty\sum_{m=l_n }^{m_n } 
\cdots , \;\; m_2=0.   
\label{ko21 } \eeqn  
As 
\beqn y''\to \left \{  
{\begin{array}{l}(z^n(\log z)^m)'\to z^{n-m\epsilon -2 }  
\;\;\left (\begin{array}{l}  
\mbox{The term such that } \\  
\mbox{$n-m\epsilon$ is the smallest} \end{array}\right )  
\;\; (^\exists a_{nm} \neq 0) \\ 
\left [\{ a_{nk_nk_{ni } }z^{k_n }(\log z)^{k_{ni} }\}  
'^2+\{ a_{nk_nk_{ni } }z^{k_n }(\log z)^{k_{ni} }\}  
'' \right ] \\ 
\hspace{49mm} 
\times e^{\sum_{i=k_n }^\infty\sum_{j=l_{ni }}^{k_{ni } } 
a_{nij }z^i(\log z)^j}\;\;\; ( ^\forall a_{nm }=0) \end{array}} 
\right.\label{ko22 } \eeqn  
for  $z\to +0$,  
\beqn \frac{y'' }{y }\to \left \{ \begin{array}{l} 
z^{n-m\epsilon -3 }\;\;  
(a_{10 }\neq 0\; \mbox{and}\;  ^\exists a_{nm}\neq 0)\\ 
z^{2k_n-2k_{ni }\epsilon -2 }e^{a_{nk_nk_{ni } }z^{k_n } 
(\log z)^{k_{ni } }}\to 0\;\;  
(a_{10 }\neq 0\; \mbox{and}\; ^\forall a_{nm }=0) \\ 
z^{-2 }\;\; (a_{10 }= 0\; \mbox{and}\;  ^\exists a_{nm }\neq 0) \\ 
z^{2k_n-2k_{ni }\epsilon-2 }\;\;(a_{10 }= 0\; \mbox{and}\; ^\forall  
a_{nm }=0)  
\end{array} \right. . \label{ko23 } \eeqn  
The possible values of $\nu_y$ for (I) remain the same: 
$\nu_y\leq -2 +\epsilon \; ; \;\; -1\leq \nu_y$. 
 
\vspace{.1 in} 
\noindent (g) Whole of the expansions obtained from type (f)  
expansions by finite times of summations, subtractions,  
multiplications, divisions (by $\neq 0$), differentiations, and  
compositions (with the shape of $f(g(z)), \; 0\leq k_g, \; g( +0)=  
+0$ where $f, g$ are type (f) expansions). 
 
This type of expansion is very complicated compared to  
an ordinary Laurent expansion, but in any case has a countable  
number of terms and a nonzero `radius of convergence' $r$  
{\footnote{Of course, the meaning is different from the usual  
one. }} 
where $y$ is analytic for $0<|z|<r$. This can also be ordered  
partially in the ascending powers and we can write the first  
term explicitly, and so monotonically diverges or converges but  
never oscillates as  $z\to +0$. Its general shape is the whole  
sum  
\beqn (1)_i +(2)_j +\cdots +(m)_k \; , \label{ko25 } \eeqn  
where 
\beqn (1)_i \;\; & := & \mbox{\huge (}\sum_{n\in \{ n\}_i 
}^\infty\sum_{m_1,\cdots , m_{d_i }=-\infty }^{m_{i1 }, \cdots , m_{id_i 
} }a_{inm_1\cdots m_{d_i } }z^n(-\log z)^{m_1 }(-\log (-z/\log 
z))^{m_2}\nonumber \\ &    &  \hspace{4.2 cm}\cdots (-\log (-z/(-\log  
(-z/\log\cdots z))))^{m_{d_i } } \mbox{\huge )}_i, \nonumber \\ 
(2)_{\pm j } & := & \sum_{i\in \{ i\}_j }(\pm )e^{\pm (1)_i },  
\nonumber \\ 
(3)_{\pm k } & := & \sum_{j\in \{ j\}_k }(\pm )e^{\pm (2)_j },  
\nonumber \\ 
\vdots \label{ko24 } \eeqn 
Here the ($\pm$) in front of $e$ takes each of the  
signatures depending on each $i$ (or $j, k, \cdots $), while the  
$\pm$ on the shoulder of $e$ and in front of $j, k, \cdots $  
takes the signature such that the coefficient of the first term  
in $\sum$ is of the same signature as $j$ after choosing the  
signatures. Each term is ordered in partially ascending powers 
with regards for any sums. The sum with index  $n$ is performed  
according to the monotonically non-decreasing sequence of real  
numbers $\{ n_i\}\; (-\infty <n_i)$ depending on $i$. In the  
same manner, the sum with index  $i, j, \cdots $ is performed  
according to the finite, monotonically non-decreasing sequence $\{  
i_j\} , \;\{ j_k\}\cdots $ of natural numbers. $m_{i_1 },  
\cdots , m_{i_{d_i } }$ take finite values, but they increase  
in correspondence with $n$ and grows $\to\infty$ as $n\to  
\infty$, and depend on $i$. $d_i$ is the maximal `depth' of the  
composition of $\log$s, or the number of $\log$s, depending on $i$  
and of finite value.  
{\footnote{The power is smaller when $m_1 +m_2 +\cdots +m_{i_1 } 
$ is greater for the same $n$, and when it is also the same and  
$m_1$ is smaller, and when it is also the same and $m_2$ is  
smaller, ..., and so on.} } 
 
As the sum of the shape of $(m)_i$ can always be represented  
as the $\exp$ of the infinite sum of the same shape,  
\beqn (m)_i & = & (\pm )e^{(m)_0 }, \;\; (m)_0:=\log \left (\mbox{sum  
of the finite number of ${e^{(m-1)_i } }$s}\right ) \nonumber \\ 
& = & (m-1)_1 +\log \left ({1\pm e^{(m-1)'_2 } +\cdots } 
\right ), \label{ko26 } \eeqn 
type (g) expansion can in fact be written in only `one term'  
$\exp (m)_{i +1 }$. 
 
Now, for the part of $i\leq 0$ in $(m)_i$, satisfying $0\leq  
k_{(m)_i }$, $\exp (m)_i$ can be written within the shape of $(m) 
_i$ as the composition of $e^z$ and $(m)_i$, Then we can write for  
(I) 
\beqn y=bz +\sum_{n=2 }^\infty a_nz^n\sim +\sum_{i<0 } 
(\pm )e^{-b_iz^i\sim\cdots }+\sum_{j<0 }(\pm )e^{-e^{c_jz^j\sim\cdots }. 
.. }  \nonumber \\ 
+\sum_{k<0 }(\pm )e^{-e^{e^{d_kz^k\sim\cdots }\cdots } 
\cdots }\cdots , \label{ko27 } \eeqn 
where $b_i, c_j, d_k, \cdots >0$, $\sim$ represents the  
abbreviation of $\log z\sim$, and $\cdots$ the higher order  
terms. The power of $y''/y$ can be classified by whether $b=0$  
or not, and what is the first of $b_i, c_j, d_k, \cdots $ such  
that the corresponding term is not 0: 
\beqn \frac{y'' }{y }\to\left \{ \begin{array}{l}(\pm )\;z^{n-m\epsilon - 
3 }\;\; (b\neq 0\;  
\mbox{and}\;  ^\exists a_n\neq 0, \; n-m\epsilon \geq 2) \\ 
(\pm )\;0\;\; (b\neq 0\; \mbox{and}\; ^\forall a_n=0\; \mbox{and}\;  
 ^\exists b_i\;\mbox{or} \; c_j\; \mbox{or}\; d_k\cdots >0) \\ 
+z^{-2 }\;\; (b=0\; \mbox{and} \;  ^\exists a_n\neq 0) \\ 
+z^{2i\pm 2\epsilon -2 }\;\; (b= ^\forall a_n=0\; \mbox{and} \;  ^\exists  
b_i>0) \\ 
+\infty\;\; (b= ^\forall a_n= ^\forall b_i=0\;  
\mbox{and} \;  ^\exists c_j\; \mbox{or} \; d_k\;  
\mbox{or} \cdots >0)\end{array} \right. ,\label{ko29 } \eeqn 
where $^\forall b_i=0$ means that there is no term in $\sum_{i<0 }$. 
 
After all, $\nu_y\leq -2 +\epsilon , \;\; -1\leq \nu_y$ for (I),  
where $\epsilon$ represents the power like $\log z\sim$. 
 
\vspace{.1 in} 
\noindent (h) It is unclear to me whether there are other cases. 
But we shall not discuss such cases further, for type (g) 
expansion is closed in usual operations, and thus is most general.
 
\vspace{.2 in} 
\noindent  (CASE 3) $y(z)$ has a non-isolated essential  
singularity at $z=0$. 
 
\vspace{.1 in} 
\noindent (i) When we allow complex coefficients in (g).  
The discussion above is almost valid in this case, except that  
when $a$ is complex $e^{az }$ shows oscillatory behavior, and  
so $y$ is not monotonic as $z\to 0$ and generally has an accumulation point  
of poles or essential singularities, keeping us away from  
defining $k_y$, $\mu_y$, or $\nu_y$. For example,  
\beqn y=z^5\sin (z^{-1 }) \eeqn  
satisfies the condition of (I) and the term with the  
smallest power in $y$ cancels that of $y''$, yet higher order  
oscillation remains.  
 
\vspace{.1 in} 
\noindent (j) It is unclear to me whether there are other cases.  
In such a case $\nu_y$ would not be clearly physical, even if defined. 
Therefore, we shall neglect such possibilities.
\section{Physical Explanation of the Result } 
The above result is not mathematically perfect, but shows that very  
wide types of functions such that closed in usual operations, 
only by satisfying the second order differential equation, 
can restrict the behavior of the potential. 
Or physically, if there exists a wave function that  
can be applied to every point of the world, the point of nonzero  
charge should also be included in the domain,  
which determines the shape of a force.
We will see this for more general case in the next section. 
 
Notice that type (g) expansion is valid under the special rule that we  
must not decompose an exponential until the end of the calculation.  
Each expansion has several infinite series of different order.  
Having nonzero `radius of convergence', it can be calculated as  
a usual function. Instead, near $z=0$, if we do not obey the  
rule and try to calculate by extracting all the terms below a certain  
order, the result, even if finite, may depend on the arrangement  
of terms. (It is known in mathematics that an infinite series that 
does not converge absolutely does not always converge to a unique value.) 
This implies an interesting non-commutative property.  
 
Notice also that the difficulties caused by point-like  
particles may be absent here. If we assume that the existence of  
an eigen function is more fundamental than that of a potential,  
there can be the region where the potential is not defined (where  
the eigen function is 0). Even if the analyticity of  
matter field is not a quantity distinguished by finite times  
of measurement, this inevitable ambiguity may be  
the origin of gauge uncertainty\cite{Simon}. 
\section{Extension to higher dimensions } 
We can extend the results to a spatial dimension $N$ as follows.  
Let us consider a spherical-symmetric Klein-Gordon equation 
with a time-independent $U(1)$ gauge potential 
$A^\mu :=(\phi (r), 0, 0, 0)$ (only the first time component is 
nonzero and the rest $N-1$ components are $0$), 
\beqn -\Delta y
& = & \frac{(E-e\phi )^2-m^2c^4}{{\hbar}^2c^2}y\nonumber \\
& =: & -V(r)y.\label{KG} \eeqn
For simplicity, we assume that 
the eigen function $y$ is a N-dimensional spherical 
symmetric function $R(r)$. For $a=0$ and $N\neq 1$,  
(\ref{ko29 }) is clearly replaced by  
\beqn \frac{\Delta R(r) }{R(r)} & = & 
 \frac{R''}{R}+\frac{N-1}{r}\frac{R'}{R} \nonumber \\ 
 & \to & \left \{ \begin{array}{l} +(N-1)r^{-2}\;\; (a\neq 0) \\ 
+n(n+N-2)r^{-2 }\;\; (a=0\; \mbox{and} \;  ^\exists a_n\neq 0) \\ 
+(-ib_i)^2r^{2i\pm 2\epsilon -2 }\;\; (a= ^\forall a_n=0\; \mbox{and} \;  ^\exists  
b_i>0) \\ 
+\infty\;\; (a= ^\forall a_n= ^\forall b_i=0\; \mbox{and} \;  ^\exists  
c_j\; \mbox{or} \; d_k\;  
\mbox{or} \cdots >0)\end{array} \right. .\label{ko30 } \eeqn 
 
We can extend the results to $r\to \infty$ case as follows. If we 
 change the variable to $z:=\frac{1}{r}$ and assume that $R(z)$ is  
$C^2$-class (expanded as below)  
\beqn R=a+bz  & + & \sum_{n=2 }^\infty  
a_nz^n\sim\cdots +\sum_{i<0 }(\pm )e^{-b_iz^i\sim\cdots } 
\cdots  \nonumber \\ 
 & + & \sum_{j<0 }(\pm )e^{-e^{c_jz^j\sim\cdots }. 
.. }\cdots +\sum_{k<0 }(\pm )e^{-e^{e^{d_kz^k\sim\cdots }\cdots } 
\cdots }\cdots , \label{ko27' } \eeqn 
(\ref{ko30 }) is clearly replaced by 
\beqn \frac{\Delta R(r) }{R(r)} & = & 
 \frac{1}{R(z)}\left \{{\frac{dz}{dr}\frac{d}{dz} 
\left ({\frac{dz}{dr}\frac{dR(z)}{dz}}\right ) 
+(N-1)z\frac{dz}{dr}\frac{dR(z)}{dz}}\right \} \nonumber \\ 
 & = & z^4\frac{R''(z)}{R(z)}-z^3(N-3)\frac{R'(z)}{R(z)} \nonumber \\ 
 & \to & \left \{ \begin{array}{l} (3-N)\frac{b}{a}z^{3}\;\; (a\neq 0\; 
  \mbox{and} \; b\neq 0\; \mbox{and} \;  N\neq 3) \\ 
(n-N+2)n\frac{a_n}{a}z^{n+2}\;\; \mbox{(or higher order)}\;\;
(a\neq 0\;  \mbox{and} \; b=0\; \mbox{and} \;  
^\exists a_n\neq 0 \; \mbox{and} \;  N\neq 3) \\ 
(n-1)n\frac{a_n}{a}z^{n+2}\;\; (a\neq 0\; \mbox{and} \; ^\exists a_n\neq 0\;  
\mbox{and} \;  N=3) \\ 
(\pm )\; 0\;\; (a\neq 0\; \mbox{and} \; b=^\forall a_n=0 \; \mbox{and} \; 
 ^\exists b_i\; \mbox{or} \; c_j\; \mbox{or} \; d_k\; \mbox{or} \cdots >0) \\ 
(3-N)z^{2}\;\; (a=0\; \mbox{and} \; b\neq 0\; \mbox{and} \; N\neq 3) \\ 
(n-1)n\frac{a_n}{b}z^{n+1}\;\; (a=0\; \mbox{and} \; b\neq 0\; \mbox{and} \;  
^\exists a_n\neq 0\; \mbox{and} \; N=3) \\ 
(\pm )\; 0\;\; (a\;  \mbox{or} \; b\neq 0\; \mbox{and} \; ^\forall a_n=0\;  
\mbox{and} \; ^\exists b_i\; \mbox{or} \; c_j\; \mbox{or} \; d_k\; \mbox{or} \cdots  
>0\; \mbox{and} \;  N=3) \\ 
(n-N+2)nz^{2}\;\; \mbox{(or higher order)}\;\; 
(a=b=0\; \mbox{and} \; ^\exists a_n\neq 0) \\ 
+(-ib_i)^2z^{2i\pm 2\epsilon +2 }\;\; (a=b= ^\forall a_n=0\; \mbox{and} \;  ^\exists  
b_i>0) \\ 
+\infty\;\; (a=b= ^\forall a_n= ^\forall b_i=0\; \mbox{and} \;  ^\exists  
c_j\; \mbox{or} \; d_k\;  
\mbox{or} \cdots >0)\end{array} \right. .\label{ko31 } \eeqn 
{\footnote{The line 2 includes the case $a\neq 0$ and $b=0$ and 
$^\exists a_n\neq 0$ and $N=n+2\neq 3$, when \\
$\frac{\Delta R(r) }{R(r)}\to (m-N+2)m\frac{a_m}{a}z^{m+2}$ or the like, where 
$a_m$ is the term next to $a_nz^n$. \\ 
The line 8 includes the case $a=b=0$ and $^\exists a_n\neq 0$ and $n=N-2$, when \\
$\frac{\Delta R(r) }{R(r)}\to (m-N+2)m\frac{a_m}{a_n}z^{m-n+2}$ or the like, where $a_m$ is the term next to $a_nz^n$.\\
In addition, line 2, 4, 7, 8 includes the Yukawa potential case, when \\
$\phi\sim Z^le^{-\frac{b_i}{z^i}}$ in (\ref{C3}), the only finite 
solutions are that of the footnote 6, i.e., 
$\frac{\Delta R(z) }{R(z)}\to \frac{d}{a}{(b_ii)}^2z^le^{-\frac{b_i}{z^i}}\;\; 
\left (R(z)=\left \{{\begin{array}{l}
a+bz^n+dz^ke^{-\frac{b_i}{z^i}}+\cdots\;\; (b=0\; \mbox{if}\; N\neq n+2)\; \mbox{or}\\
az^n+dz^{k+n}e^{-\frac{b_i}{z^i}}+\cdots\;\; (N=n+2)
\end{array} }\right.\right )$, where 
$a, d\neq 0$ and $b_i, i>0$ and $k=2i-l$. It is curious that there are some 
`degenerate' eigen functions for the same asymptotic potential, 
even if not normalizable for $N<4$.\\
Finally, this and line 9 allow for a single term with pure imaginary $b_i$, 
except for which an imaginary coefficient on the shoulder of the exponential 
leads to a non-physical oscillatory or imaginary potential. This is indeed 
the case for the Coulomb scattering of a photon.
}}
Noting that $2\leq n$ and $i<0$, we conclude the potential 
 $V(r)$ as $r\to\infty$ must be positive for ($N\leq 3$ and 
$\nu =-2$) or $-2< \nu$, where $\nu$ is the power of the potential 
 $V\to r^\nu$ as $r\to\infty$; can take both signs for 
other cases.
There is no reason to assume that $R(z)$ is $C^2$-class, but more
 natural normalizability condition that $R(r)$ is a $L^2$ function leads
 to small modification $N<2n$ instead of $2\leq n$ in (\ref{ko27' }) 
and so, $a=0$ if $0<N$ and R.H.S. of (\ref{ko31 }) is replaced by
\footnote{It is impossible for the R.H.S. to be
\beqn \to \left \{ \begin{array}{l} 
(n-1)n\frac{a_n}{b}z^{n+1}\;\; (a=0\; \mbox{and} \; b\neq 0\; \mbox{and} \;  
^\exists a_n\neq 0\; \mbox{and} \; 1<n\; \mbox{and} \; N=3) \\ 
(\pm )\; 0\;\; (a=0\;  \mbox{and} \; b\neq 0\; \mbox{and} \; ^\forall a_n=0\;  
\mbox{and} \; ^\exists b_i\; \mbox{or} \; c_j\; \mbox{or} \; d_k\; 
\mbox{or} \cdots >0\; \mbox{and} \;  N=3) \\ 
\end{array} \right. .\nonumber \eeqn 
because $b$ appears. In addition, `higher order' does not appear in case of 
$n\leq 2$ nor $N\leq 4$.\\
For the realistic $N=3$ case, the weaker $L^2$ condition to allow logarithmic 
divergence is equivalent to $R(z)$ of $C^1$-class, with only difference 
that $1<n$ instead of $2\leq n$ in (\ref{ko30 }) and (\ref{ko30 }).}
\beqn \to  \left \{ \begin{array}{l} 
(n-N+2)nz^{2}\;\; \mbox{(or higher order)}\;\; 
\left ({\begin{array}{l}
(a=b=0\; \mbox{and} \; ^\exists a_n\neq 0\; \mbox{and} \; N<2n)\; \mbox{or}\\
(a=0\;\mbox{and} \; ^\exists a_n\neq 0\;\mbox{and} \; \frac{N}{2}<n<1)
\end{array} }\right )\\ 
+(-ib_i)^2z^{2i\pm 2\epsilon +2 }\;\; (a=b= ^\forall a_n=0\; \mbox{and} \;  ^\exists  
b_i>0) \\ 
+\infty\;\; (a=b= ^\forall a_n= ^\forall b_i=0\; \mbox{and} \;  ^\exists  
c_j\; \mbox{or} \; d_k\;  
\mbox{or} \cdots >0)\end{array} \right. .\label{ko32 } \eeqn 
 In this case, the potential must be 
positive for ($\nu =-2$ and $N<n+2$) or $-2<\nu$.
 Notice that (\ref{ko30 }) for more
general cases of $N, a$ can be obtained from (\ref{ko31 }) by the trivial 
replacement $N\to 4-N$ and $z\to r$ with its power smaller by $4$. 
Then, we conclude the potential $V(r)$ as $r\to +0$ must be 
positive for ($1\leq N$ and $\nu =-2$) or $\nu < -2$, where $\nu$ is the power of the potential $V\to r^\nu$ as $r\to +0$; can take both signs for 
other cases. If we assume $R(r)$ is $L^2$ instead of $C^2$, 
$-2n<N$ instead of $2\leq n$, and so by renaming $a_1:=b$ the results are
\beqn & \frac{\Delta R(r) }{R(r)} & \nonumber \\
 & \to & \left \{ \begin{array}{l} 
(n+N-2)n\frac{a_n}{a}r^{n-2}\;\; \mbox{(or higher order)}\;\;
(a\neq 0\; \mbox{and} \; ^\exists a_n\neq 0\; \mbox{and} \;  
0<n) \\ 
(\pm )\; 0\;\; (a\;\mbox{or}\; a_{2-N}\neq 0\;  \mbox{and} \; ^\forall a_n=0\; 
\; \mbox{for} \;n\neq 2-N\; \mbox{and} \; ^\exists b_i\; \mbox{or} \; c_j\; 
\mbox{or} \; d_k\; \mbox{or} \cdots  >0) \\ 
(n+N-2)nr^{-2}\;\; \mbox{(or higher order)}\;\; 
\left ({\begin{array}{l}
(a=0\; \mbox{and} \; ^\exists a_n\neq 0)\; \mbox{or}\\
(^\exists a_n\neq 0\;\mbox{and} \; n<0)
\end{array} }\right ) \\ 
+(-ib_i)^2r^{2i\pm 2\epsilon -2 }\;\; (a=^\forall a_n=0\; \mbox{and} \; 
^\exists  b_i>0) \\
+\infty\;\; (a= ^\forall a_n= ^\forall b_i=0\; \mbox{and} \;  ^\exists  
c_j\; \mbox{or} \; d_k\;  
\mbox{or} \cdots >0)\end{array} \right. .\label{ko34 } \eeqn 
\footnote{The line 1,3 includes the special case $N=2-n$, when 
$\frac{\Delta R(r) }{R(r)}\to$
$(m+N-2)m\frac{a_m}{a}r^{m-2}$, $(m+N-2)m\frac{a_m}{a_n}r^{m-n-2}$ 
or the like, respectively, where $a_m$ is the term next to $a_nr^n$
such that $m\neq 0$.
In addition, `higher order' does not appear in case of $n\leq -2$ 
nor $4\leq N$.}
Above results show that for a physical dimension $N=1, 2, 3$,
the sign of a potential $V$ must be positive for
$\nu\leq -2+\epsilon\;\; (r\to 0)$ and
$-2-\epsilon\leq\nu\;\; (r\to\infty)$, but can be negative for other cases.

\section{Theorems for the long distance limit} 
Now we define the following conditions for later convenience. \\
The first are normalization conditions naturally required for the 
long and distance limit of a boson or fermion free field. Clearly, 
{\bf Normalization  Conditions for Free fields} are \\
for {\bf Massive Boson field in the Long distance limit (NC-MBL)},\\
$n\leq -\frac{N+1}{2}$, \\
for {\bf Massless Boson field in the Long distance limit (NC-0BL)},\\
$n\leq \frac{1-N}{2}$, \\
for {\bf Massive Fermion field in the Long distance limit (NC-MFL)},\\
$n\leq -\frac{N+1}{2}$, \\
for {\bf Massless Fermion field in the Long distance limit (NC-0FL)},\\
$n\leq -\frac{N}{2}$, \\
for {\bf Boson field in the Short distance limit (NC-BS)},\\
$\frac{1-N}{2}\leq n$, \\
for {\bf Fermion field in the Short distance limit (NC-FS)},\\
$-\frac{N}{2}\leq n$, \\
where the field behaves like $r^n$ and $=$ means logarithmic divergence, \\
all of them with the {\bf Exceptional rule for massless particles (NC-0Ex)} 
that \\
an arbitrary constant can be added. 
\\ 
It's unclear to me  whether or not previous results are valid for the Dirac 
equation, but for a moment we keep away the validity or spin effects 
as a later discussion, and just describe the specific results given by 
application for fermions by writing in a [ ].\\
{\bf Positive Potential Condition (PPC)} \\ 
The potential $V(r)$ defined in section 2 is positive.\\
 This is indeed satisfied for the non-relativistic approximation of 
a Klein-Gordon equation (\ref{KG}), if 
\beqn -V & = & \frac{(E-e\phi )^2-m^2c^4}{{\hbar}^2c^2} \label{C1} \\
& \approx & \frac{2m}{\hbar^2}(E-mc^2-e\phi )<0,\label{C2}\eeqn
where $|E-mc^2|, |e\phi |\ll mc^2$. In addition, (\ref{C1}) shows that 
PPC is strictly valid for massless bosons. Indeed, PPC means that 
the particle is in the bound state. A force is defined as 
$-e\nabla\phi$. Thus from the previous result we 
can verify the following theorems for the long distance limit.\\
{\bf Theorem 1} \\ 
For the higher or smaller spatial dimension $N\neq 3$, a massless boson [or fermion] 
that behaves like $\sim \frac{1}r\;\; (r\to\infty )$ can not feel a dominant 
$\frac{1}{r^2}$-like long range force. \\
{\bf Proof} \\ 
For $N<3$, this is proven by taking $E=m=0$ in (\ref{C1}) and comparing with 
the line 5 of (\ref{ko31 }). The former violates PPC, which contradicts the 
latter condition that $V(+\infty)$ must be positive for $N<3$. For $3<N$, 
this is proven just because nonzero $b$ in the line 5 breaks {\bf NC-0BL} 
and for $3<N$, that is a weaker condition than {\bf NC-MBL}, {\bf NC-0FL}, 
{\bf NC-0FL}.\\
\\
 A static spherical symmetric electric field like $\sim \frac{1}{r^2}$ is 
of course experimentally observed and therefore, a photon behaves like 
$A^\mu =(\phi (r), 0, 0, 0),\;\;\phi(r)\sim\frac{1}{r}$. An interesting corollary
of the above theorem is that, if $N>3$, a photon can not feel gravity 
and there is no gravitational lens! Some people might suspect that we can not 
always take $E=0$ because it means a virtual photon for $e\phi >0$, but 
at least in situation a photon is bounded by the potential and another 
photon is not bounded, we can always take $E=0$ and thus we dare say \\
{\bf Theorem 2} \\ 
For $N>3$, if a charged static spherical symmetric black hole can exist, 
the electric field decreases more rapidly than $\frac{1}{r^{N-1}}$. \\
{\bf Proof} \\ 
Even a black hole has its gravitational potential $\sim\frac{1}{r}$ 
at distance, for its density is finite and spherical symmetric, 
and obeys Gauss' law and $\frac{1}{r^2}$-law experimentally, 
and gravity is always attractive force.  
A black hole is of course such a matter that even a light can 
not escape, therefore a bound state exist for a photon. 
Then, we can apply theorem 1. Strictly speaking, there might be exceptions for 
the theorem, in which the black hole potential is deviated from 
the $\frac{1}{r}$-law because of the presence of another long range 
force that a photon can feel. In the standard model of particles, 
this is not the case, for no other long range force 
(gluons nor a photon) couple with a photon. Noting that 
if the asymptotic $\frac{1}{r^2}$-law of gravity holds, from (\ref{KG}), 
\beqn 
V(r) & = & \frac{m^2c^4-E^2+2eE\phi -(e\phi )^2}{{\hbar}^2c^2} \nonumber \\
& \to &  \left \{ \begin{array}{l} 
(1)\;\;m^2c^4-E^2\;\; (E^2\neq m^2c^4) \nonumber \\ 
(2)\;\;2eE\phi\sim \frac{1}{r}\;\; (E^2=m^2c^4\neq 0) \nonumber \\ 
(3)\;\;(e\phi )^2\sim \frac{1}{r^2}\;\; (E^2=m^2c^4=0) 
\end{array} \right. .\label{C3} \eeqn 
Thus the only ways for the massless photon to allow such a black hole is 
the lines 8,9 of (\ref{ko31 })
{\footnote{
For $
N<3$ (and also for $N=3$ if we do not allow logarithmic divergence), 
this can also be derived from {\bf NC-0BF} without assuming 
the $\frac{1}{r^2}$-law of gravity. For [fermions and] massive bosons, 
more severe than logarithmic divergence appears.}
}
i.e., 
\begin{description}
	\item[(line 8.)] 
This is (3) of (\ref{C3}), where the asymptotic $\frac{1}{r^2}$-law 
of gravity exactly holds and the photon remains massless, but 
the electric field behave as if away from a polarized matter with 
no electric charge as a whole. The former requires that no 
density is present at distance. Gauss' law is geometric and 
valid in presence of gravity, therefore the latter requires 
real existence of the charge to cancel that of the black hole. 
Up to now all the particles with electric charge are massive, 
therefore the cancellation must be due to the electric charge 
density distributed in a finite region. From (\ref{C3}) 
and {\bf NC-0BL} and (\ref{ko31 }), such a `medium range' force 
can be felt dominant only by such fields that behave like 
($b=0$ and $\frac{N-1}{2}\leq n$). (For other particles, 
Notice that in quantum mechanics, even a particle in a empty 
metal sphere can `feel the outer world'. 
Or
	\item[(line 9.)] 
The electric field vanish exponentially, and $V(r)$ survive slowly 
than $\frac{1}{r^2}$. For each case of (\ref{C3}), 
(1) $i=-1$ (2) $i=-\frac12$ (3) no $i$ allowed.
Thus only possible cases are, either the photon `becomes massive' 
(i.e., $m\neq 0$) or otherwise $E\neq 0$. It is an interesting possibility 
that a massive static photon, that is not bounded because violating PPC, 
can create $e^{\sqrt r}$-type electric field 
in the former case and even a massless photon can create 
Yukawa-type electric field in the latter case. 
Normalization condition is automatically satisfied for these 
exponentially vanishing solutions. Such a Yukawa-type electric field 
can be felt iff by a $b=0$ massless boson [or fermion] satisfying the
normalization condition. 
\\
 There is yet another possibility that asymptotic $\frac{1}{r^2}$-law 
of gravity changes to survive more slowly, because of 
the long tail of nonzero density the black hole is accompanied with. 
In this case, $i$ can take some negative value $\neq -1$ 
iff ($E^2=m^2c^4$ and $-1< i<0$) or $i<-1$, when the photon create neither 
long range nor Yukawa-type but rapidly vanishing electric field. \\
In addition, from (\ref{ko31 }), this is the only case for gluons 
to make $\phi\sim r$ potential at distance, when $i=-2$ regardless of 
$m, E$.
\end{description}
{\bf Theorem 3} \\ 
For $N=3$, a massless boson [or fermion] vanishes more rapidly than  
$\sim \frac{1}{r}\;\; (r\to\infty )$, if it feels a 
$\frac{1}{r^2}$-like dominant long range force. \\
{\bf Proof} \\ 
This is proven by taking $m=0$ in (\ref{C3}) and comparing with the 
line 8, 9 of (\ref{ko31 }), for they are the only cases for 
$\phi\sim\frac{1}{r}\;\; (r\to\infty)$ to exist. \\
\\
Thus, Theorem 2 and its proof hold also for $N=3$, only by adding 
the last of {(\bf line 8)} $<$and {(\bf line 9)}$>$ the following sentences:
`except for the line 6 $<$ and 7$>$ of (\ref{ko31 }), in which only 
logarithmic divergence appears for a massless boson that feels the photon 
which behaves like $b\neq 0$. But it causes a self contradiction 
to identify the massless boson as a photon feeling itself' . \\
An interesting corollary of Theorem 1, 3 is \\
{\bf Corollary 1} \\ 
A massless gauge boson that feels a dominant 
$\frac{1}{r^2}$-like long range force can not create a long range force. \\ 
\\ 
This is a bit strange, for a photon can not feel $\frac{1}{r}$-like 
potential of gravitational lens. Maybe $\frac{1}{r^2}$-rule 
of electric field is only approximation in presence of gravity, 
or $\frac{1}{r^2}$-rule of gravity is only approximation 
in presence of electric field, or the coexistence of a graviton 
and a photon leads to a contradiction in present theory and 
gravity should be derived from other forces. 
But this corollary well accounts for the properties of the 
standard model, for if a gluon or a glueball had an 
electric charge, it must be `massive', and if a photon had a color, 
the photon must be `massive', provided an isolated gluon or a glueball 
could be observed,  
In addition, a weak boson has an electric charge (this is followed 
by the experimental fact that an electron is suddenly created and 
comes out in $\beta$-decay), and then it must be `massive', 
regardless of the Higgs mechanism. Thus we come to \\
{\bf Corollary 2} \\ 
If a photon or graviton is massless and the self interactions of 
$W^\pm$ bosons are not so strong as to create a long range force 
which vanish more slowly than $\frac{1}{r^2}$, then the 
$W^\pm$ bosons can not create a long range force. In the same way, 
if a graviton is massless, the glue-balls (if exist) and pions 
with weak enough self interaction can not create a long range force, 
and even if a graviton is massive and a photon is massless, the 
electrically charged glueballs (if exist) and pions with weak 
enough self interaction can not.\\
If the standard model particles are to be unified some day in 
such a manner that a photon is massless and a graviton has an 
electric charge, then the graviton must be 'massive'. 
In the same way, if a graviton is massless, then the photon 
must be `massive'. Conversely, 
if a graviton is massless and a photon or gluon is massive, 
then the photon or gluon must be `massive' (this is tautology). \\
\\
By the way, from the argument of footnote 6, we can verify also \\
{\bf Theorem 4} \\ 
If a boson [or fermion] feels a dominant Yukawa-type potential, 
then the eigen function of the particle is not $L^2$. Particularly, 
such a boson [or fermion] must be massless for $N<5$; 
can be massive for $5\leq N$. 
\\
{\bf proof} \\ 
This is because if we take $E=mc^2$ and Yukawa-type potential 
$\phi\sim Z^le^{-\frac{b_i}{z^i}}$ in (\ref{C3}), the only finite 
solutions are that of the footnote 6, i.e., 
$\frac{\Delta R(z) }{R(z)}\to 
\frac{d}{a}{(b_ii)}^2z^le^{-\frac{b_i}{z^i}}\;\; 
\left (R(z)=\left \{{\begin{array}{l}
a+bz^n+dz^ke^{-\frac{b_i}{z^i}}+\cdots\;\; (b=0\; \mbox{if}\; N\neq n+2)\; \mbox{or}\\
az^n+dz^{k+n}e^{-\frac{b_i}{z^i}}+\cdots\;\; (N=n+2)
\end{array} }\right.\right )$, where 
$a, d\neq 0$ and $b_i, i>0$ and $k=2i-l$, 
which vanish no more rapidly than $r^{2-N}$, to be compared 
with {\bf NC-MBL}, {\bf NC-0BL}, [{\bf NC-MFL}, {\bf NC-0FL}], 
and {\bf NC-0Ex}. Notice that for $N=3$, a logarithmic divergence inevitably 
appears even for massless boson in the lower case, and therefore this case 
is impossible for $3<N$. \\
It is surely a severe condition for realistic physics and thus \\
{\bf Corollary 3} \\ 
A short range force must be always dominated by a longer range force, 
or otherwise $N<5$ and not felt by a massive boson [and fermion], 
or otherwise $5\leq N$, \\
where the term `longer' used to notice that any force that survives 
more slowly than any Yukawa potential is allowed. \\
\\
From now on, we take $N=3$ and concentrate on the self-consistent 
conditions for the standard model. Then, if we do not take account of 
gravity as the dominant force, \\
{\bf Corollary 4} \\ 
A boson with a charge of a short range force must have another 
charge of a longer range force, or otherwise must be massless. \\
\\ 
Thus, the unification of $U(1)\times SU(2)$ can be proved without 
assuming experimental results. The corollary well accounts 
for the property of $W^\pm$ [and quarks and charged leptons] 
which are massive and have both $U(1)$ and $SU(2)$ charge. 
An equivalent proposition that 
`a particle with no charge of any long range force must not have 
a charge of any short range force, or otherwise must be massless' 
is satisfied for a $Z^0$ and a photon and a $\pi^0$ 
[and almost for neutrinos]. Only if we can neglect gravity$\cdots$.
\section{Origin of Higgs mechanism}
Now, remember that a massless free spin $1$ boson is always 
identified with a photon and creates a $\frac{1}{r^2}$-like 
long range force which is identified with electric field, 
and gravity couples equivalently to all matter 
\cite{WeinG}. Suppose that a massive boson has a charge. If we 
do not take account of gravity as the dominant force, and 
if it is the charge of a short range force, then from 
{\bf Corollary 4} the boson has a charge of a longer range force. 
Thus, we can naturally assume that a `boson with a charge of 
a short range force' has a charge of a long range force, 
say, electric charge, and call it $W^+$, for a photon of 
course exists and creates a long range force. 
Then, from the $CPT$-theorem, its anti-particle $W^-$ also 
exists. And this is the source of the short range force, 
therefore must be identified with the `photon that feels 
a $\frac{1}{r^2}$-force and can create a Yukawa-type electric field' 
previously occurred in the case (1) of (\ref{C3}), for the Gauss' law 
is also valid for the field of the `reshaped photon' and 
gives the real charge distribution. Notice that in this case, 
there is no way to distinguish whether or not the `reshaped photon' 
is massive, for the $V(r)$ becomes the constant $m^2c^4-E^2$ 
asymptotically, and a massless photon with a positive energy $E$ 
is equivalent to an energy-less photon with the mass $mc^2:=iE$. 
Thus any photon to feel the same asymptotic nonzero $V(\infty )$ 
can create the same Yukawa-type electric field and therefore can be 
taken to be massless. Indeed, the Klein-Gordon equation 
for a massless photon with the 
$\frac{1}{r}$-potential $\phi_G$ (of gravity) can be solved 
from (\ref{ko31 }) 
to give 
\beqn 
\frac{\Delta R(r) }{R(r)} & = & -\frac{(E_\gamma -\phi_G )^2}{{\hbar}^2c^2} 
\nonumber \\ 
& \sim & m^2-2mn\frac{1}{r}+n(n-1)\frac{1}{r^2} \nonumber \\ 
& = & -\frac{(E_W-\phi_G-e\phi_{EM})^2-{M_W}^2c^4}{{\hbar}^2c^2}, 
\nonumber \\
R(r) & \sim & e^{-mr}r^n \label{C7} \eeqn 
This is in fact the definition of the mass, i.e., 
for a massless and charge-less photon to feel gravity, 
some universal unit for the mass is needed. Here we define 
the unit by the fine structure constant $\alpha :=\frac{e^2}{\hbar c}$. 
Then, we can write $\phi_G:=\frac{e^2 {\cal M}}{r}$, where $\cal M$ 
is some constant proportional to the mass the photon feels. 
In the same way, the electric potential that the $W^+$ feels 
can be written as $\phi_{EM}:=\frac{e{\cal Q}}{r}$, where $\cal Q$ 
is some constant proportional to the electric charge the $W^+$ feels. 
Thus, the initial energy $E_\gamma$ of the massless photon feeling the
`universal potential of gravity' $\phi_G$ 
is equal to the potential that the identified $W^+$ boson 
feels, i.e., asymptotically ${E_\gamma}^2={E_W}^2-{M}_W^2c^4$ 
with $E_W$ and $M_W$ the energy and mass of the $W^+$ boson respectively. 
From the {\bf line 9} of (\ref{ko31 }) with $i=-1$, 
$E_W$ is equal to ${b_{-1}}^2=:-m^2$, which 
in turn creates `reshaped photon' potential $\phi_W \sim\frac{e^{-mr}}{r}$.
In the low energy limit of the photon $E_\gamma =0$, this means that a 
`stopped photon' is just a static electric field. 

{\bf This is the origin of the Higgs mechanism.} 
Therefore, from (\ref{C7}) 
\beqn 
\begin{array}{lll}
{\left \{ \begin{array}{l}
\alpha ({\cal M+Q}) E_W =-mn\hbar c\hspace{14mm}\\
(m\hbar c)^2=-{E_\gamma}^2={M}_W^2c^4-{E_W}^2\end{array}\right.} & 
\to & 
{\left \{\begin{array}{l}
-{E_\gamma}^2=(m\hbar c)^2\\
=\frac{{M}_W^2c^4}{(1+\frac{n^2}{{e'}^2})},
\end{array}\right.}\hspace{-49mm}
\end{array} \nonumber \\
\mbox{with } {e'}^2:=\{\alpha({\cal M+Q})\}^2 
\label{G7} 
\eeqn
\\
for $W^\pm$. In the same way, 
\beqn 
\begin{array}{lll}
{\left \{ \begin{array}{l}
\alpha ({\cal M}) E_Z =-mn\hbar c\hspace{14mm}\\
(m\hbar c)^2=-{E_\gamma}^2={M}_Z^2c^4-{E_Z}^2\end{array}\right.} & 
\to & 
{\left \{\begin{array}{l}
-{E_\gamma}^2=(m\hbar c)^2\\
=\frac{{M}_Z^2c^4}{(1+\frac{n^2}{{e''}^2})},
\end{array}\right.}\hspace{-49mm}
\end{array} \nonumber \\
\mbox{with } {e''}^2:=\{\alpha({\cal M})\}^2 
\label{G8} 
\eeqn
\\
for $Z^0$. But, wait, the mass of $W^\pm$ must always be pure imaginary! 
Something is wrong$\cdots$.
\section{Unification}
Let's keep away the problem of imaginary mass, and consider the original 
eigen function of the photon. The $\frac{1}{r^2}$-rule of gravity may 
not be altered much, for it is always attractive. Then, from 
{\bf Theorem 3} the photon can not create an exactly $\frac{1}{r^2}$-like 
long range force. But it can create {\bf almost $\frac{1}{r^2}$-like} 
medium range force, by taking $n=1+\epsilon$ in the line $8$ 
of (\ref{ko31 }), without violating {\bf (NC-0BL)}. Maybe a radical, but 
not so contradictory solution is just to expand a most general eigen 
function $R(z)$ in a shape like (\ref{ko27' })
\beqn R(z)=a+bz  & + & \sum_{n=1 }^\infty  
a_nz^n\sim\cdots +\sum_{i<0 }(\pm )e^{-b_iz^i\sim\cdots } 
\cdots  \nonumber \\ 
 & + & \sum_{j<0 }(\pm )e^{-e^{c_jz^j\sim\cdots }. 
.. }\cdots +\sum_{k<0 }(\pm )e^{-e^{e^{d_kz^k\sim\cdots }\cdots } 
\cdots }\cdots , \label{U1} \eeqn 
{\footnote{The only deference is that the first infinite sum can start 
from $z^{1+\epsilon}$, which means to assume $C^1$-class $\phi (z)$ 
instead of $C^2$, but causes no problem for a moment.}}
and assume the shape of a graviton $G(z)$ as the first infinite sum 
part of (\ref{U1}) with $a=0\neq b$ and no $i, j, k,\cdots$. 
Then, we can always take $0<b$ for a graviton without losing 
generality, and redefine the gravity potential by $\phi_G:=g_GG(z)$, 
where $g_G$ is the positive coupling constant of gravity. 
Then, the gravity becomes {\it by definition} attractive. 
Let's consider a virtual world in which only gravitons exist. 
Because $\phi_G$ is universal to any particles, it must also be 
satisfied for the graviton. Therefore, 
\beqn\frac{\Delta G(z) }{G(z)} & = & -\frac{(E_G -\phi_G )^2}
{{\hbar}^2c^2}.  \label{U2} \eeqn 
This is a self-consistent condition that resembles the Einstein 
equation in a sense, but from {\bf Theorem 3} there is no solution 
for (\ref{U2}) to create a $\frac{1}{r^2}$-like long range force, 
and (\ref{U2}) leads to $b=0$. This indicates that a graviton must be 
accompanied with other field, say photon. From above discussion, 
we can naturally assume the shape of a photon $A(z)$ as the first 
infinite sum part of (\ref{U1}) with $a=b=0$ (may start from 
$z^{1+\epsilon}$) and no $i, j, k,\cdots$. 
Then, we can redefine the electric potential and gravity potential 
by $\phi_{EM}:=g_{EM}A(z)$ and $\phi_{G}:=g_{EM}bz$, 
where $g_{EM}$ is the positive coupling constant of electricity. 
But in this case, there is no way to 
make $\phi_{EM}$ always positive, as it depends on the relative 
sign of $a_n$ to $0<b$ in $G(z)$. With these redefinition, 
a graviton and a photon suit well for (\ref{ko31 }) and (\ref{C7}).
Let's abandon (\ref{U2}) and interpret $G(z)$ as a {\it virtual} 
field of the first term $bz$. Then, the only equation for $A(z)$ to satisfy is 
\beqn\frac{\Delta A(z) }{A(z)} & = & -\frac{(E_\gamma -\phi_G )^2}
{{\hbar}^2c^2}, \nonumber \\
\mbox{where } \phi_G & = & g_{EM}bz. \label{U3} \eeqn 
This has a consistent solution for $E_\gamma =0$ as in the 
{\bf (line 8)} of (\ref{ko31 }). 
{\footnote{(\ref{U1}) is closed in this shape of iterative 
expansion and no $i, j, k, \cdots$ appears.}}
\\
Let's come back again to ({\ref{C7}}). Exactly $\frac{1}{r}$-like 
potential comes only from the first $g_{EM}\frac{b}{r}$ term in this equation. 
For $W^\pm$ to make a short 
range force, the solution $W^\pm (z)$ must be the {\bf (line 9)} case of 
(\ref{ko31 }). For almost all the unbounded states, $Mc^2<E_W$. Then, from 
the previous discussion we must take 
$i=-1, |m|=b_i=E_\gamma=\sqrt{{E_W}^2-M^2c^4}>0$ in 
the {\bf (line 9)}. From footnote $6$ this `$W^\pm$ boson with imaginary mass' 
is just a free photon. Notice that $E_\gamma$ can always be taken 
positive {\it by definition} and negative energy means just a complex 
conjugate. Let's denote these eigen functions of unbounded photons 
(those were not included in the previous expansion because of imaginary 
coefficients) as $A^\pm(z)$, and previous one (with $0$ energy) $A^0(z)$. 
Contrastingly, for a bounded state $E_W<Mc^2$, 
$i=-1, m=|b_i|=|E_\gamma |=\sqrt{{M^2c^4}-{E_W}^2}>0$ and this $W^\pm$ boson 
is really massive. But for $M$ to be real, only pure imaginary $E_\gamma$ 
is allowed, and $m$ must be positive to satisfy normalization conditions. 
With this $i$, $W^\pm (z)$ is identified to be the second infinite 
sum appears in the general expansion (\ref{U1})
{\footnote{This expansion is also closed in itself if we take $i=-1$, 
in such a sense that no other $i$ appears in the iterative expansion 
$\displaystyle W^\pm(z)=\sum_{k=1,2,3\cdots }^\infty\sum_{l_k} 
e^{-k\frac{m}{z}}z^{l_k}$, except if we consider the special case 
$E_W=Mc^2$ when a half integer $i=-\frac12$ must also be included.
If we must assume (\ref{U1}) to be Taylor expanded in $r$, we can avoid 
other eccentric exponential decay with a non-integer $i=-1$ or appearance of 
$j,k,\cdots$. }}
. Experimentally, $M_W<M_Z$. Then from (\ref{G7}) and (\ref{G8}) 
${\cal Q}<-2{\cal M}$ or $0<{\cal Q}$, the latter of which 
is in good accordance with the interpretation that gravity is a 
virtual force induced by electric polarization. Suppose that 
there are two metal balls, one is neutral and the other with positive 
or negative electric charge. Then, in both case the two balls will 
attract each another by the induced surface charge. Then, what is 
the meaning of ${\cal M}$ and ${\cal Q}$ ? $\cdots$ well, say, 
the mass and the charge induced on the surface of the universe, 
for they are universal constants and every matter is bounded by gravity 
in the universe. Isn't it an interesting idea? \\ 
Then, let's consider the eigenfunction $W^\pm (z)$ and $Z^0(z)$. The only 
difference of them are that $W^\pm$ can feel electricity but $Z^0$ 
can not. Therefore in a theory that includes no graviton like the standard 
model, from {\bf Theorem 4} a massive $Z^0$ boson must be 
written as $Z^0(z)=b_Zz+d_Zz^{1-n''}e^{-\frac{m}{z}}+\cdots$, where 
$b_Z, d_Z\neq 0$, while $W^\pm (z)=\sum_{n=1}^\infty {a_W}_nz^n\sim +\cdots 
+d_W\sum_{k=1,2,3\cdots}^\infty z^{1-n'}e^{-k\frac{m}{z}}+\cdots$, where 
${a_W}_n, d_W\neq 0$ and $n', n''$ are the $n$ s which satisfy (\ref{G7}) and 
(\ref{G8}) respectively. But from (\ref{C7}) this is a virtual shape of 
a massless photon only feeling gravity and with energy 
$E_\gamma :=\pm im\hbar c$. 
Above discussion is valid only if $E_\gamma$ can take a specific 
pure imaginary value. This is artificially accomplished by the redefinition of 
$\phi_G$, i.e., taking $G(z)\to G(z)\pm i\frac{m\hbar c}{g_E}$ in
(\ref{U3}). 
In the standard model, neutrino mass must exactly be $0$ because their eigen 
functions must be 
\eq \nu (z)=a_\nu +d_\nu z^{-n''}e^{-\frac{m}{z}}+\cdots\label{nu}
\en 
to avoid divergence. 
Strictly speaking, $Z^0(z)$ must also have this shape, i.e., 
\beqn Z^0(z) & = & a_Z +d_Z z^{-n}e^{-\frac{m}{z}}+\cdots \nonumber \\ 
(b=^\forall a_n & = & 0 \;\;\mbox{because it can not feel electric field) 
and then }
\label{Z} \nonumber \\
W^\pm(z) & = & a_W +\sum_{j=1}^\infty {a_W}_jz^j\sim +\cdots 
+d_W z^{-n}e^{-\frac{m}{z}}+\cdots\nonumber \\
(^\exists{a_W}_j & \neq 0 & \;\;\mbox{because it can feel electric field), } 
\label{W} \nonumber \\
\mbox{where} & n & \mbox{is the solution of (\ref{G7}) with }
\phi_G=g_{EM}bz)
\nonumber \eeqn 
to avoid logarithmic divergence and thus must be massless. 
The only way to allow massive $Z^0$ is then to subtract $a_Z$ from every particles that feel weak force by using the constant ambiguity of a potential. 
This requires Higgs vacuum energy $<v>\neq 0$. Then, for $Z^0$ 
and $W^\pm$ to feel weak force of the same strength, 
$d_W=d_Z$ and we must replace $-n\to 1-n$ in the term 
$z^{-n}e^{-\frac{m}{z}}+\cdots$. Then, at last gauge bosons [and neutrinos 
and charged leptons (and quarks)] in the 
standard model can be unified (or decomposed) in the following form:
{\footnote{
Usage of these equations: If you want to calculate the two-body 
problem of $X$ and $Y$, use the reduced mass $\frac{M_XM_Y}{M_X+M_Y}$ 
for the last term of a potential $V$, and the particles obey the 
Klein-Gordon equation such that both particles feel the same
potential. Then, the eigen function for this system can be obtained 
from $X(z)$, $Y(z)$ such that both expansions contain the same order terms.
Notice that previous discussions and symmetry requirement lead to 
$b \sim {\cal M} \sim M_X+M_Y$, when the Newtonian potential and 
its Schwarzshild correction can naturally be derived by scaling $E$ by 
the unit $\frac{M_XM_Y}{2(M_X+M_Y)}c^2$.
$A^0(z)$ must start from just $n=n_{min}=2$ for a massive $W^\pm$ to stop.
${a_W}_n$ and ${a_L}_n$ and ${a_Q}_n$ must start from at least 
$2+\epsilon$ to avoid logarithmic divergence.}}
\beqn 
R(z) & = & A^0(z)+A^\pm (z)+Z^0(z)+W^\pm (z)+\nu (z)+L^\pm (z)(+G(z)+Q^\pm (z)),\;\;
\mbox{where} \nonumber \\
A^0(z) & := & \sum_{n=n_{min}}^\infty {a_A}_nz^n\sim\cdots\;\;\mbox{and}
 \nonumber \\
A^\pm(z) & := & \sum_{k=1,2,3\cdots }^\infty\sum_{l_k=1}^\infty
{d_A}_{kl_k}^\pm e^{\pm ik\frac{\omega}{z}}z^{l_k}\;\;\mbox{satisfy} 
\nonumber \\
\frac{\Delta A(z) }{A(z)} & = & -\frac{(\pm E_\gamma -\phi_{G})^2}
{{\hbar}^2c^2} \;\;\mbox{with} \nonumber \\
E_\gamma & = & 
{\left \{ \begin{array}{l}
im\hbar c\;\; (\mbox{for}A^0(z))\\
(im+\omega )\hbar c\;\;\mbox{, where }\omega\mbox{ is any positive number}\;\; 
(\mbox{for}A^\pm(z))\\
\end{array}\right.}, \nonumber \\
Z^0(z) & := & (a_Z-bz)\left (\delta_{E_ZM_Z}\pm\sum_{k=1,2,3\cdots }^\infty
\sum_{l_k=1}^\infty 
{d_Z}_{kl_k} e^{-k\frac{m}{z}}z^{l_k}+\cdots\right )\;\;\mbox{and}\nonumber \\
\nu^0(z) & :\approx & a_\nu\left (\delta_{E_\nu M_\nu}\pm\sum_{k=1,2,3\cdots }^\infty
\sum_{l_k=1}^\infty {d_\nu}_{kl_k}e^{-k\frac{m}{z}}z^{l_k}+\cdots\right )
\;\;\mbox{and}\nonumber \\
W^\pm (z) & := & a_W\left (\left (\pm 1\;\;\mbox{or}\;\;\sum_{n=2}^\infty 
{a_W}_nz^n\sim\right )\delta_{E_WM_W}+\sum_{k=1,2,3\cdots }^\infty
\sum_{l_k=1}^\infty {d_W}_{kl_k} e^{-k\frac{m}{z}}z^{l_k}+\cdots\right )
\;\;\mbox{and}\nonumber \\
L^\pm (z) & :\approx & a_L\left (\left (\pm 1\;\;\mbox{or}
\;\;\sum_{n=2}^\infty 
{a_L}_nz^n\sim\right )\delta_{E_LM_L}+\sum_{k=1,2,3\cdots }^\infty
\sum_{l_k=1}^\infty {d_L}_{kl_k} e^{-k\frac{m}{z}}z^{l_k}
+\cdots\right )\;\;\mbox{and}\nonumber \\
Q^\pm (z) & :\approx & a_Q\left (\left (\pm 1\;\;\mbox{or}\;\;
\sum_{n=2}^\infty {a_Q}_nz^n\sim\right )\delta_{E_QM_Q}+
\sum_{k=1,2,3\cdots }^\infty\sum_{l_k=1}^\infty 
{d_Q}_{kl_k} e^{-k\frac{m}{z}}z^{l_k}+\cdots\right )\nonumber \\
 &   & \mbox{or}\;\;
\sum_{k'=1,2,3\cdots }^\infty \sum_{l'_{k'}=1}^\infty {f_Q}_{k'l'_{k'}} 
e^{-k'\frac{M}{z^2}}z^{l'_{k'}}+\cdots\;\;\mbox{satisfy} \nonumber \\
\frac{\Delta Z^0(z) }{Z^0(z)} & = & -\frac{(\pm E_Z-\phi_{G}
-\vec{I_Z}\cdot\vec{\phi_W})^2-{M_Z}^2c^4}
{{\hbar}^2c^2}\;\;\mbox{and}\nonumber \\
\frac{\Delta \nu(z) }{\nu(z)} & \approx & -\frac{(\pm E_\nu -\phi_{G}
-\vec{I_\nu}\cdot\vec{\phi_W})^2-{M_\nu}^2c^4}
{{\hbar}^2c^2}\;\;\mbox{and}\nonumber \\
\frac{\Delta W^\pm (z) }{W^\pm (z)} & = & 
-\frac{(\pm E_W-\phi_{G}-e\phi_{EM}-\vec{I_W}\cdot\vec{\phi_W})^2-{M_W}^2c^4}{{\hbar}^2c^2}
\;\;\mbox{and}\nonumber \\
\frac{\Delta L^\pm (z) }{L^\pm (z)} & \approx & 
-\frac{(\pm E_L-\phi_{G}-e\phi_{EM}-\vec{I_L}\cdot\vec{\phi_W})^2-{M_L}^2c^4}{{\hbar}^2c^2}
\;\;\mbox{and}\nonumber \\
\frac{\Delta Q^\pm (z) }{Q^\pm (z)} & \approx & 
-\frac{(\pm E_Q-\phi_{G}-q\phi_{EM}-\vec{I_Q}\cdot\vec{\phi_W}\pm C_G\phi_S)^2-{M_Q}^2c^4}{{\hbar}^2c^2}, 
\mbox{where}\nonumber \\
\phi_{G} & := & g_{EM}bz+im\hbar c\;\;\mbox{and}\nonumber \\
\phi_{EM} & := & g_{EM}A^0(z)\;\;\mbox{and}\nonumber \\
\vec{\phi_W} & := & g_{EM}
(W^+(z)\delta_{E_WM_W}-a_W, W^-(z)\delta_{E_WM_W}-a_W, 
Z^0(z)\delta_{E_ZM_Z}-a_Z)\;\;\mbox{and}\nonumber \\
\vec{d} & := & ({d_{W^+}}_{11}, {d_{W^-}}_{11}, {d_Z}_{11}),\;\;
{I^3}_Xa_Zg_{EM}:=im\hbar c, \nonumber \\
{M_Z}^2c^4-{E_Z}^2 & = & (m\hbar c)^2 = {M_W}^2-{E_W}^2\;\;\mbox{and}
\nonumber \\
M_Z\frac{\vec I_Z\cdot\vec d}{d_Z}_{11} & = & \frac{(m\hbar
c)^2}{2g_{EM}} 
= M_W\frac{\vec I_W\cdot\vec d}{d_W}_{11}
\approx M_\nu\frac{\vec I_\nu\cdot\vec d}{d_{\nu}}_{11}
\approx M_L\frac{\vec I_L\cdot\vec d}{d_L}_{11}
\approx M_Q\frac{\vec I_Q\cdot\vec d}{d_Q}_{11}
\;\;\mbox{and}\nonumber \\ 
\frac{n(n-1)}{eM_W} & = & 2\frac{g_{EM}}{\hbar^2 c^2}{a_A}_2\approx 
\frac{n(n-1)}{eM_L}\approx\frac{n(n-1)}{qM_Q},\;\;\mbox{where}\nonumber \\
{a_W}_nz^n & \sim & \;\;\mbox{is the first term for the stopped} 
\;\;W^\pm ,\;\;\mbox{etc.,}\nonumber \\ 
G^\pm (z) & := &  \sum_{n=-1}^\infty {a_G}_nz^n\sim +\cdots 
+\sum_{k'=1,2,3\cdots }^\infty \sum_{l'_{k'}=1}^\infty 
{f_G}_{k'l'_{k'}}e^{-k'\frac{M}{z^2}}z^{l'_{k'}}\sim +\cdots
\;\;\mbox{satisfies} \nonumber \\
\frac{\Delta G^\pm (z)}{G^\pm (z)} & = & -\frac
{(\pm E_G-\phi_G-{\vec{C}_G}\cdot\vec{\phi_S})^2}{\hbar^2c^2}, 
\;\;\mbox{where}\nonumber \\
\vec\phi_{S} & := & g_{S}\vec G(z), \;\; 
-\left (\frac{g_s\vec C_G\cdot\vec{a_G}_{-1}}{\hbar c}
\right )^2=(2M)^2, \;\; -(\frac{g_{EM}b}{\hbar c})^2=n_{min}(n_{min}-1).
\eeqn
Thus, the God made the light at first, or a man can, by defining all the 
coupling constants after $g_{EM}\cdots$.
\section{Conclusion} 
In this paper we classified possible singularities of a potential for 
the spherical symmetric Klein-Gordon equation, assuming that a  
potential $V$ has at least one $C^2$-class eigen function. The result  
crucially depends on the analytic property of the eigen function near its  
0 point. Above analysis indicates that possible shapes of the potential 
and the eigen functions of particles are restricted by the 
consistency condition of this simple model. Then we discussed
a natural possibility that gravity and weak coupling constants 
$g_G$ and $g_W$ are defined after $g_{EM}$. 
In this point of view, gravity and the weak force are subsidiary derived 
from electricity. 
The fact that the iterative solution inevitably includes several  
infinite series with different order in one expansion may be 
the origin of the non-commutative gauge invariance. 
 
\section*{Acknowledgments} 
I am grateful to Izumi Tsutsui and Toyohiro Tsurumaru for  
useful discussion. This paper is partially motivated by 
some implication given by Tsutomu Kambe and Kazuo Fujikawa. 
I also appreciate Syu Kato and my family Yuko, Takeo, 
Megumi, and Masahide for spiritual support. 
This work is partly supported by Ikuei-kai grant.
Finally, I thank Maya Y for her respectable life. 
 
\end{document}